%% file: WearLevelingTechniqueFor_NVM_Caches.tex
\documentclass[conference,onecolumn]{IEEEtran}

\usepackage{graphicx}

\usepackage{multirow}
\usepackage{epsfig}
\usepackage[square, comma, sort&compress,numbers]{natbib}
\usepackage{amssymb} 

\makeatletter
\def\Ginclude@eps#1{%
 \message{<#1>}%
  \bgroup
  \def\@tempa{!}%
  \dimen@\Gin@req@width
  \dimen@ii.1bp%
  \divide\dimen@\dimen@ii
  \@tempdima\Gin@req@height
  \divide\@tempdima\dimen@ii
    \includegraphics{#1}%
  \egroup}
\makeatother

\usepackage{wrapfig}
\usepackage{booktabs}

\usepackage{amsmath}
\usepackage{relsize}

\newcommand\blfootnote[1]{%
  \begingroup
  \renewcommand\thefootnote{}\footnote{#1}%
  \addtocounter{footnote}{-1}%
  \endgroup
}

\usepackage[linesnumbered,lined,ruled,commentsnumbered]{algorithm2e}
\usepackage{times}

\usepackage[table]{xcolor}  
\usepackage{color}

\usepackage{url}

\makeatletter
\def\url@leostyle{%
  \@ifundefined{selectfont}{\def\UrlFont{\sf}}{\def\UrlFont{\small\ttfamily}}}
\makeatother
\begin{document}
\bstctlcite{IEEEexample:BSTcontrol}
\title{ Using Cache-coloring to Mitigate Inter-set Write Variation in Non-volatile Caches}

 \author{\IEEEauthorblockN{Sparsh Mittal }
 \IEEEauthorblockA{Department of Electrical and Computer Engineering \\
 Iowa State University, Ames, Iowa 50011, USA\\
  Email: sparsh0mittal@gmail.com}
}


\maketitle

\input{abstract}



%
\IEEEpeerreviewmaketitle

\input{intro}

\input{background}

 \input{methods}
 
 \input{experimental_platform}
 \input{results}

\input{conclusion}



\ifCLASSOPTIONcaptionsoff
  \newpage
\fi



\bibliographystyle{IEEEtran}
\bibliography{PhDReferences}

\end{document}

%% file: abstract.tex
\begin{abstract}
In recent years, researchers have explored use of non-volatile devices such as STT-RAM (spin torque transfer RAM) for designing on-chip caches, since they provide high density and consume low leakage power. A common limitation of all non-volatile devices is their limited write endurance. Further, since existing cache management policies are write-variation unaware, excessive writes to a few blocks may lead to a quick failure of the whole cache. We propose an architectural technique for wear-leveling of non-volatile last level caches (LLCs). Our technique uses cache-coloring approach which adds a software-controlled mapping layer between groups of physical pages and cache sets. Periodically the mapping is altered to ensure that write-traffic can be spread uniformly to different sets of the cache to achieve wear-leveling. Simulations performed with an x86-64 simulator and SPEC2006 benchmarks show that our technique reduces the worst-case writes to cache blocks and thus improves the cache lifetime by 4.07$\times$.
\end{abstract}

%% file: intro.tex
\section{Introduction}

 To bridge the gap between the speed of processor and main memory, modern processors are using large on-chip caches \cite{intelxeoncache}. The design of modern processors is becoming increasingly constrained by power consumption \cite{Mit_DRAMsurvey} and hence, due to several limitations of SRAM, such as high leakage power consumption, low density and poor scaling, their use for designing large last level caches (LLCs) in exascale performance systems has become challenging. Researchers have recently explored use of alternative non-volatile devices such as STT-RAM (spin torque transfer RAM), PCM (phase change memory), ReRAM (resistive RAM) \cite{jog2012cache,wang2013i2wap} etc. These devices provide high density and low leakage power which make them an attractive choice for design of large LLCs. 
 
 A common limitation of non-volatile memories, however, is that their write endurance values are much smaller than those of SRAM and DRAM \cite{qureshi2011phase}. Further, since the existing cache management schemes are not write variation-aware; due to the write non-uniformity, heavily-written cache blocks can fail much earlier than most other blocks of the cache. Due to this phenomenon, not only a cache sensitive application, but even a cache insensitive application may also cause the endurance limit to be reached in a few days. For this reason, use of wear-leveling techniques is extremely important for managing wearable caches. \blfootnote{Sparsh Mittal is currently working as a postdoctoral research associate at Oak Ridge National Laboratory, USA. }

 In this report, we present a wear-leveling technique for  LLCs designed with non-volatile memory devices. Our technique uses cache coloring scheme to add a software-controlled mapping layer between groups of physical pages (called memory regions) and cache sets. Periodically, our technique computes the number of writes to different colors of the cache and based on it, the mapping of a few colors is changed to channel the write traffic to least utilized cache colors. This helps in  achieving wear-leveling. In cache, both inter-set and intra-set write non-uniformity may occur. Our technique works to mitigate inter-set write non-uniformity and can be synergistically integrated with other techniques for mitigating intra-set write non-uniformity (e.g. \cite{wang2013i2wap}).

Our technique has several salient features. It performs cache set remapping at coarse granularity of cache color level  and not at set level. As an example, a 16-way, 4MB cache has only 64 colors but 4096 sets. Thus, using coarse granularity leads to smaller data collection and algorithm computation overhead. The length of an interval is decided by the number of writes and hence, our technique easily adapts itself based on the write-intensive nature of a program. Further, our technique uses software control with lightweight hardware support, and thus can be easily implemented in real-processors.  Finally, while we show the working of our technique for STT-RAM caches in this work, our technique is also applicable to other non-volatile devices.

Simulations have been performed using Sniper x86-64 simulator \cite{CarHei2011_Sniper} and SPEC2006 benchmarks. The results show that our technique significantly reduces the number of worst-case writes to the cache set, which leads to increase in the cache lifetime. On average, the relative lifetime of cache on using our technique is 4.20$\times$. Also, our technique has small implementation overhead and by virtue of using dynamic  profiling (and not offline profiling), it is suitable for use in product systems.

The remainder of the report is organized as follows. Section \ref{sec:endurancebackground} provides a brief background on emerging memories and existing wear-leveling techniques. Section \ref{sec:endurancemethods} presents our methodology and the algorithm for wear-leveling.  Section \ref{sec:enduranceexerimentalmethodology} presents the simulation platform. Section \ref{sec:enduranceresults} presents the experimental results and finally, Section \ref{sec:enduranceconclusion} provides the conclusion.

%% file: background.tex
\section{Background and Related Work}\label{sec:endurancebackground}
\subsection{Non-volatile Memories and Need of Wear-leveling Techniques}
Processors have conventionally used SRAM for designing on-chip caches. Since their leakage power consumption is high,  several architecture-level techniques have been proposed to reduce the leakage energy dissipation \cite{MitZha13_Cashier,KaxHuz01_CacheDecay,mittal2013cachebook}. However, future exascale computing systems present much higher demands of performance and energy-efficiency and hence, the  low density and poor scalability of SRAM present bottleneck in its use for future exascale computing systems. This has forced the researchers to explore alternative devices.   Embedded DRAM (eDRAM) is a volatile device which provides better density and lower leakage power than SRAM, however, its retention period is in the range of micro-seconds \cite{barth2008500} and hence, it may spend a large amount of energy in the form of refresh energy \cite{mittalEDRAM2013}.

Among non-volatile devices, flash memory is nearly 200$\times$ slower than DRAM and has a write-endurance of only $10^4$-$10^5$ writes. For this reason, it is typically used as a solid state disk (SSD) and not as cache or memory \cite{qureshi2009enhancing}. Phase change memory (PCM) is only $2\times$-$4\times$ slower than DRAM and has $4\times$ more density than DRAM, however, its write endurance value is much smaller than that of SRAM and DRAM \cite{mittalPhaseChange2013}.  For spin torque transfer RAM (STT-RAM), although the write-endurance has been estimated to be $10^{15}$, the best endurance test result so far shows a write endurance of only $4\times 10^{12}$ writes \cite{huai2008spin}. Also, its high write energy and power need to be addressed for enabling its use as a cache. For resistive RAM (ReRAM), the write endurance is $10^{11}$ \cite{kim2011bi}. Also, it has been shown \cite{sheu20114mb}, that ReRAM can provide $15\times$ better density than SRAM and a read/write latency which is comparable to an SRAM cache of similar capacity. Although its write access energy is $10\times$ that of SRAM, its low-leakage property leads to significant reduction in the energy consumption. Thus, with a suitable wear-leveling technique, NVMs can be used for designing on-chip caches.

\subsection{Wear-leveling Techniques for Caches}
It is well-known that the number of accesses to different sets of the cache are non-uniform \cite{rolan2009adaptive}. This leads to inter-set write variation. This effect is especially significant for non-volatile caches, since their write endurance is small. Recently, researchers have proposed techniques for addressing write-variation in NVM caches. The existing wear-leveling techniques for main memory (e.g. \cite{qureshi2009enhancing,6513577}) may not be directly applied to address write-variation in cache. Caches possess several unique properties and by taking advantage of these features, more effective wear-leveling techniques can be designed for caches. 

Wang et al. \cite{wang2013i2wap} present techniques for addressing intra-set and inter-set write-variation in ReRAM based last level caches. Their technique for addressing inter-set write-variation periodically swaps the cache sets in order to distribute the write-traffic uniformly. Their technique for addressing intra-set write-variation flushes a block which sees write-hit, after a fixed number of writes. Since there is high probability that this would be a hot line, their technique deliberately adjusts the write traffic to different ways of the cache reduce intra-set write variation. Chen et al. \cite{Chen2013OCB} propose a technique to address inter-set write-variation in STT-RAM based last level caches. Their technique uses a register called remap register. At the end of each interval, their technique updates this register and uses XOR between this register and the set-index bit of cache address. Using this, the set-index of all addresses in the cache is changed.

Syu  et al. \cite{syu2013high} propose a hybrid cache design that includes SRAM banks, STT-RAM banks, and STT-RAM/SRAM hybrid banks. For a 4-core CMP with 16 banks, 4 banks are designed as SRAM banks, another 4 as STT-RAM/SRAM hybrid banks and the remaining 8 banks are designed as STT-RAM banks. In the hybrid bank, 15 out of 16 ways are designed with STT-RAM. By redirecting write-traffic to SRAM, the write pressure on STT-RAM can be reduced and the performance can be improved. For wear-leveling of STT-RAM, they provision that when a write request has cache hit of STT-RAM line and there is an invalid SRAM line, the write request is redirected to the invalid SRAM line and the STT-RAM line is invalidated. The SRAM lines in hybrid bank receive the write request from the STT-RAM line of the same hybrid bank only. To make the write pressure among different cache partitions as uniform as possible, they propose a partition-level wear leveling scheme which considers the different workload write-pressure from each core to change the capacity of SRAM and STT-RAM regions in each partition.

%% file: methods.tex
\section{Methodology}\label{sec:endurancemethods}
  In this section, we present the key idea and working of our technique.

\subsection{Key Idea}

The main idea of our approach is that physical addresses which are heavily (frequently) written should be periodically mapped to different sets of the cache, so that the number of writes to a single (or few) set(s) does not increase much more than the average value. For this, we first need to choose a suitable granularity of cache remapping. It can be a single set (as used in \cite{Chen2013OCB}), however, this incurs large profiling overhead. In this work, we choose a granularity of cache color, which has large number of sets (e.g. 64 sets). In what follows, we explain our method in detail.

\subsection{Cache Coloring Scheme}
To enable flexible mapping of physical addresses to cache sets, we use cache coloring scheme \cite{mittal2013PhDThesis,KesHil92_PageColoring,MitZha13_Cashier}. It works as follows. First, the cache is logically divided into $N$ portions, called cache colors. The number of colors $N$, can be computed as follows.

\begin{equation} \label{eq:Nvalue}
N = \dfrac{CacheSize}{ PageSize \times CacheAssociativity} \end{equation}
 
Second, the physical pages are divided into $N$ \textit{memory regions} based on the LSBs (least significant bits)  of their physical page number. Cache coloring works by mapping a memory region (and hence, all physical pages in that region) to a unique color in the cache. To record this mapping,  a small  mapping table is used which stores the cache color assigned to each memory region. By changing the mapping between the physical pages and the cache colors, a particular memory region can be mapped to desired cache color. Thus, cache can be reconfigured at the granularity of a single cache color. This cache coloring scheme does  not require a change in underlying virtual address to physical address mapping, and thus can be implemented with little overhead. Also, the size of mapping table is extremely small and hence, its overhead is also small.


\subsection{ Algorithm}\label{sec:energysavingalgo}

We now provide the details of our algorithm which can be a kernel module. The algorithm runs, after every $K$ writes, if at least 3M cycles have elapsed since the last algorithm execution. This is done to ensure that the algorithm does not run very frequently for applications with very high write counts. If 3M cycles have not elapsed, the algorithm waits till 3M cycles have elapsed, and checks for the possible reconfiguration after next $K$ writes. The symbols  used in the description of the algorithm are shown in Table \ref{tab:symbols}. 

\begin{table*}[htbp]
  \centering
  \caption{Symbols used in the algorithm}
    \begin{tabular}{|l|l|}
    \hline
    Symbol & Meaning \\\hline
    \hline
    Region[$i$] & Region number which is mapped to color $i$ \\\hline
    nWriteGlobal[$i$] & Writes to color $i$ till now \\\hline
    nWriteLastInterval[$i$] & Writes to color $i$ in last interval \\\hline
    AVG & Average of nWriteLastInterval[$i$]   for all colors\\\hline
    SDW & Standard deviation of nWriteLastInterval[i]  for all colors \\\hline
    nHigher & Number of colors which have nWriteLastInterval[$i$] $>$ AVG \\\hline
    $\lambda$ & A threshold ($\le$ N/2) \\\hline
    $\beta$ & Another threshold \\\hline
    L1, L2 & Two lists \\\hline
    
    \end{tabular}%
  \label{tab:symbols}%
\end{table*}%

Let MAX(A,B) denote the function which returns the maximum value out of A and B. Also, if (r1, c1) and (r2, c2) be the region-to-color mapping for two colors c1 and c2, then we define \textit{swapping} their region-to-color mapping as obtaining a mapping (r1, c2) and (r2, c1). If c1 = c2, swapping function does not perform any action. The steps of the algorithm are as follows.

\begin{enumerate}
\item  If SDW $<$ $\beta$, the variation in write-count is small. Return.
\item Maintain two lists.  In one list L1, sort the colors based on decreasing value of nWriteLastInterval[i]. In another list L2, sort the colors based on increasing value of nWriteGlobal[i].
\item Let nColorToSwap = MAX(nHigher,$\lambda$)
\item For all $k$, 1 $\le$ $k$ $\le$ nColorToSwap 

\{

Swap the region-to-color mapping of color at location L1[k] and L2[k].

\}
\end{enumerate}
 
The algorithm works as follows. If the variation in the write counts in the last interval are small, remapping is not performed. In our experiments, we take $\beta$ = 75. The algorithm attempts to reduce the write-pressure on those colors which were heavily used in  the last interval. Through periodic remapping, the algorithm tries to channel this write-traffic to those colors which have, till now, experienced the least number of writes. For this, two lists L1 and L2 are maintained. L1 is sorted based on writes in the last interval and L2 is sorted based on accumulated writes till now. The value of  nColorToSwap decides the number of maximum colors to be swapped in an interval. $\lambda$ shows the fixed upper limit on nColorToSwap and it is taken as N/4 in our experiments. Also nHigher shows the number of colors which have higher write count than the average. Intuitively, reducing the write traffic of a color is expected to be useful only if the writes to it are higher than the average. When the mapping of a region to color is changed, all the blocks in that color are flushed, i.e. the dirty blocks are written back to memory and the clean blocks are discarded.

\subsection{Features and Limitations of the Algorithm} 
  
Previous techniques which perform cache remapping at set-level \cite{Chen2013OCB,wang2013i2wap}, do not collect per-set statistics, since that would incur large overhead. Thus, they blindly remap the whole cache periodically, which causes large flushing overhead, resulting in performance penalty. Our technique performs the mapping at coarse level to reduce the overhead, but this allows us to collect the information on cache access  with low-overhead.   By changing the value of $\lambda$, $\beta$ and algorithm interval, a trade-off can be achieved between performance loss and improvement in endurance. Since the algorithm runs after a large interval, its overhead is easily amortized over the interval length. Unlike \cite{Chen2013OCB}, our technique uses number of writes to determine the interval length and not the number of cycles. Using this, our technique easily adapts itself based on the write-intensive nature of a program.  Further, in \cite{Chen2013OCB}, due to remapping, the set index bits must be included as part of the tag. This changes the set-matching. This is avoided in our technique.
  
The limitation of our technique is that it our technique performs management only at level of cache color, which has several sets. For our experiments, a single cache color has 64 sets, and it may be possible that a single set may have large number of writes. However, this fine-grain information is lost in the coarse grain migration. 


 %

%% file: experimental_platform.tex

\section{Experimental Methodology}\label{sec:enduranceexerimentalmethodology}
\subsection{Simulation Platform}

We use interval core model in Sniper x86-64 simulator which has been verified against real-world processor \cite{CarHei2011_Sniper}. For all caches, the block size is 64B and LRU (least recently used) replacement policy is used. Processor frequency is 2GHz. L1 I/D  is 4-way set-associative cache of 32 KB size, with a latency of 1 cycle. The latency of main memory is 160 cycles and memory queue contention is also modeled. Memory bandwidth is 12 GB/s.

STT-RAM is a non-volatile device which provides the ability to trade off non-volatility to gain performance by improving write speed. In literature, different   methods have been proposed to achieve this \cite{smullen2011relaxing,jog2012cache}. In our experiments, we assume that a suitable value of retention time has been achieved by the method proposed by Jog et al. \cite{jog2012cache}. We choose a retention time of 1 second, which, at 2GHz frequency, provides a write latency value of 12 cycles. This provides significant improvement over a write latency value of 22 cycles achieved for a retention time of 10 years \cite{jog2012cache} and also avoids the refresh overhead which may be incurred in using a retention time of 10 milli-second \cite{jog2012cache}. We use a unified L2 cache which has 4MB size with 16-way set-associativity.  Its parameters are shown in Table \ref{tab:sttramenergy}.

We use all 29 SPEC CPU2006 benchmarks with \textit{ref} inputs. In the figures in the result section, for the sake of clarity, we use first three-letters of the name of benchmark as the acronyms of the benchmark. These acronyms are shown in Table \ref{tab:acronyms}. We fast-forward the benchmarks for 10B instructions and then simulate them for 400M instructions.

\begin{table}[htbp]
  \centering
  \caption{SPEC2006 Benchmarks and Their Acronyms}
    \begin{tabular}{|c|c||c|c|}
    \hline
    astar & Ast   & bwaves & Bwa \\    
    bzip2 & Bzi   & cactusADM & Cac \\
    calculix & Cal   & dealII & Dea \\
    gamess & Gam   & gcc   & Gcc \\
    gemsFDTD & Gem   & gobmk & Gob \\
    gromacs & Gro   & h264ref & H26 \\
    hmmer & Hmm   & lbm   & Lbm \\
    leslie3d & Les   & libquantum & Lib \\
    mcf   & Mcf   & milc  & Mil \\
    namd  & Nam   & omnetpp & Omn \\
    perlbench & Per   & povray & Pov \\
    sjeng & Sje   & soplex & Sop \\
    sphinx & Sph   & tonto & Ton \\
    wrf   & Wrf   & xalancbmk & Xal \\
    zeusmp & Zeu   &       &  \\
    \hline
    \end{tabular}%
  \label{tab:acronyms}%
\end{table}%

\subsection{Evaluation Metrics}
To quantify the effect of our technique, we define relative improvement in cache lifetime. The lifetime of a device such as cache or memory can be defined as either raw lifetime (determined by the first failed block) or error-tolerant lifetime (determined by raw lifetime and error recovery methods). In this work, we focus on raw lifetime and assume that error correction methods can be integrated with our technique to improve the lifetime even further, since these methods are orthogonal to our technique. 

The raw lifetime is defined as the inverse of the maximum number of writes on any cache block.  Note that although we perform mapping at the level of \textit{colors}, we take the maximum value of number of writes to any \textit{block}. This is because of the intuitive concept of lifetime. 
 
 In addition, we  show the results on relative performance (also called  speedup), percentage energy saved and absolute increase in miss-per kilo instructions (MPKI). We account for the energy consumption of L2 cache and main memory, since other components of processor are minimally affected by our technique. The energy values for L2 cache are shown in Table \ref{tab:sttramenergy}.

\begin{table}[htbp]
  \centering
  \caption{Parameters of STT-RAM 4MB (1 sec) cache \cite{jog2012cache}}
    \begin{tabular}{|c|c|c|c|c|} 
\hline	
Read Latency &	Write Latency&	Read Energy	&Write Energy &	Leakage Power \\ \hline
0.973 ns&	5.571 ns&	1.015 nJ&	1.036 nJ	&2235 mW\\\hline

    \end{tabular}%
  \label{tab:sttramenergy}%
\end{table}%

The dynamic energy of each main memory access is 70 nJ  and leakage power consumption of main memory is 0.18 Watt \cite{mittal2013cachebook,ZheLin09_DIMM}. The average value of relative lifetime  and relative performance are computed as geometric mean of per-workload values. On the other hand, the average value of energy saving and increase in MPKI are computed as arithmetic mean of per-workload values.


%% file: results.tex

\begin{figure*}[htbp]
 \centering
 \includegraphics [scale=0.5] {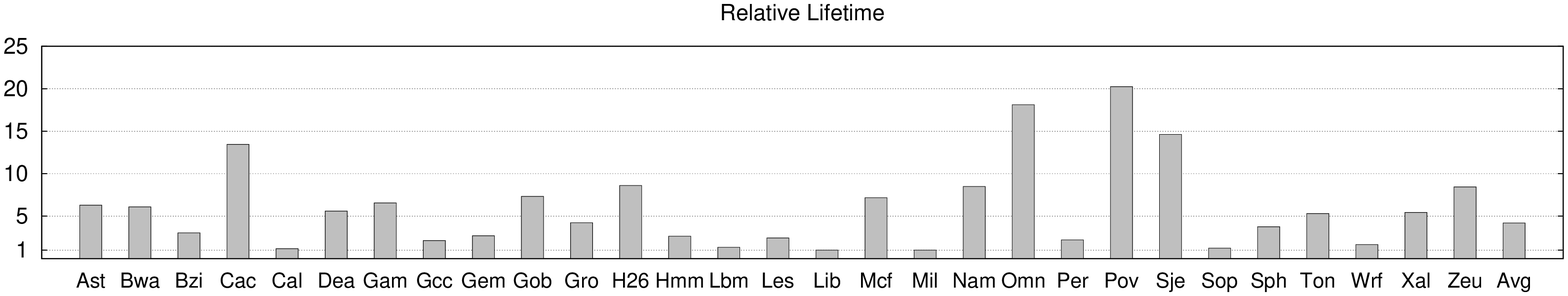}
 \includegraphics [scale=0.5] {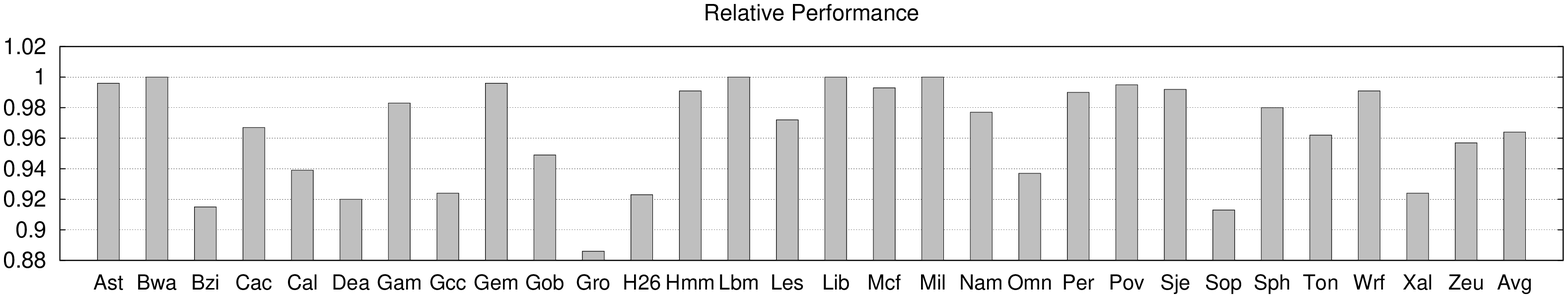}
 \includegraphics [scale=0.5] {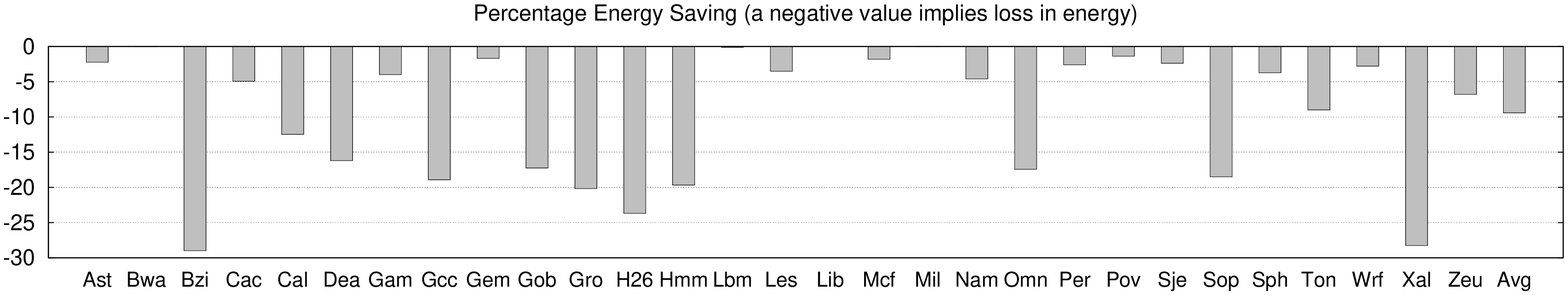}
 \includegraphics [scale=0.5] {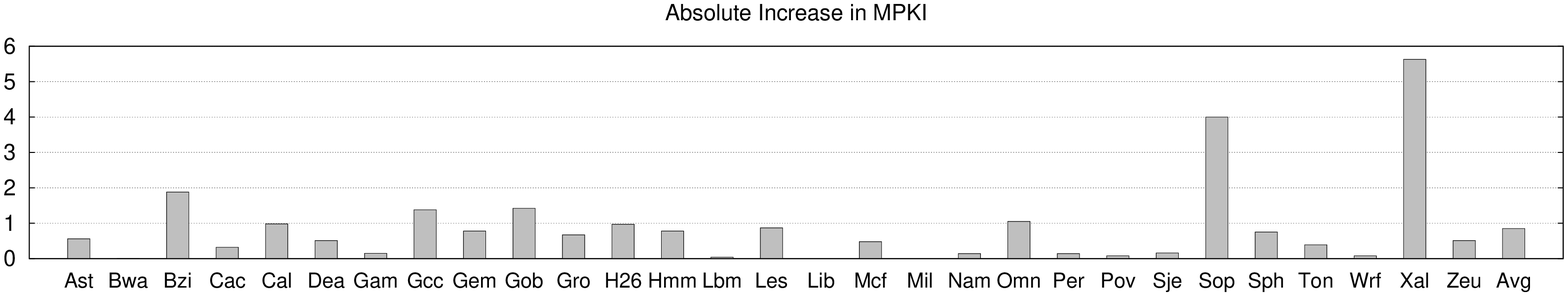}
 \caption{Results on relative lifetime, relative performance, energy saving and absolute increase in MPKI}
\label{fig:mainresults}
 \end{figure*}
 
 \section{Results}\label{sec:enduranceresults}
Figure \ref{fig:mainresults} shows the results on cache lifetime, performance, energy saving and increase in MPKI. The average improvement in cache lifetime is 4.20$\times$ and average improvement in performance is 0.96$\times$. The average loss (i.e. increase) in energy is 9.42\% and average increase in MPKI is 0.85.

From the results, it is clear that our technique improves the lifetime to more than 4$\times$ the original value. This shows the effectiveness of our approach. The average value of relative performance is 0.96, thus the average performance loss of our technique is small. Also, since the caches built with non-volatile devices typically provide better energy efficiency than conventional SRAM caches, a small loss in energy due to use of our technique may be acceptable, since our technique addresses a crucial bottleneck of non-volatile caches, which is its limited write endurance. 

Our technique can make more intelligent decisions by taking into account other performance metrics such as miss-rate, number of accesses etc. Using this, the algorithm can adapt itself to the workloads to minimize the performance and energy loss. This is planned as the future work.

It is noteworthy that if the writes to different sets are already uniform, then wear-leveling techniques cannot provide improve the lifetime. Thus, the write variation present in the original application presents the limit of the improvement that can be achieved. To show this, we show the standard deviation of write accesses to different cache blocks in the cache for different applications. Side by side, we show the relative lifetime value achieved by using our technique. 
\begin{table}[htbp]
  \centering
  \caption{Standard deviation of write accesses to different cache blocks (SD) and relative lifetime achieved using our technique}
    \begin{tabular}{|c|c|c||c|c|c|}
    \hline
    \multirow{2}{*}{Workload}  & \multirow{2}{*}{SD}   & Relative & \multirow{2}{*}{Workload}  & \multirow{2}{*}{SD}   & Relative \\    
     &  & Lifetime &  &   & Lifetime \\\hline
    
    Ast   & 91.6  & 6.3   & Lib   & 0.6   & 1.0 \\ 
    Bwa   & 54.6  & 6.1   & Mcf   & 148.7 & 7.2 \\
    Bzi   & 116.8 & 3.0   & Mil   & 227.1 & 1.0 \\
    Cac   & 20.2  & 13.5  & Nam   & 182.8 & 8.5 \\
    Cal   & 96.4  & 1.2   & Omn   & 49.0  & 18.1 \\
    Dea   & 59.5  & 5.6   & Per   & 39.2  & 2.2 \\
    Gam   & 643.6 & 6.6   & Pov   & 1163.4 & 20.3 \\
    Gcc   & 167.6 & 2.1   & Sje   & 222.4 & 14.6 \\
    Gem   & 172.7 & 2.7   & Sop   & 67.2  & 1.2 \\
    Gob   & 293.2 & 7.3   & Sph   & 27.9  & 3.8 \\
    Gro   & 143.5 & 4.2   & Ton   & 145.7 & 5.3 \\
    H26   & 50.0  & 8.6   & Wrf   & 30.5  & 1.7 \\
    Hmm   & 82.5  & 2.6   & Xal   & 426.3 & 5.4 \\
    Lbm   & 91.0  & 1.3   & Zeu   & 771.1 & 8.4 \\
    Les   & 282.3 & 2.4   &       &       &  \\
    \hline
    \end{tabular}%
  \label{tab:writevariation}%
\end{table}%

From Table \ref{tab:writevariation}, it is clear that, in general, for applications with high SD value, the relative lifetime achieved is also higher. Also note that caches have both inter-set and intra-set write variation and our technique only addresses inter-set write variation. 

For benchmarks such as libquantum etc., the SD with baseline experiments is already very small and hence, the scope for wear-leveling is small. For such applications, improvement in the lifetime needs to obtained using other methods, such as reducing the number of writes etc. For benchmarks such as povray, sjeng etc., the value of SD is very large, showing the need of a wear-leveling technique to improve cache lifetime.

Also note that for the same area, non-volatile device has much larger capacity than that of SRAM. Thus, the same number of writes are distributed to larger number of sets in a cache designed with non-volatile device. As an example, replacing a 1MB SRAM cache with 32MB ReRAM (or other NVM) of same area distributes the writes to 32$\times$ the number of sets. This extra capacity helps in intrinsically distributing the writes to larger number of sets and reducing the write pressure to each cache block. Finally, our technique can be synergistically integrated with other techniques which minimize the write traffic to cache (e.g. \cite{zhou2009energy}) and/or  mitigate intra-set write non-uniformity (e.g. \cite{wang2013i2wap}) to increase the lifetime of non-volatile caches even further.


%% file: conclusion.tex
\section{Conclusion}\label{sec:enduranceconclusion}
Integration of emerging non-volatile devices in cache hierarchy is an exciting, yet challenging research topic and several issues need to be resolved before making this a tangible reality. In this report, we presented a technique for addressing inter-set write-variation in last level cache designed with STT-RAM. Since all non-volatile devices have limited endurance, our technique is also useful for caches designed with other non-volatile devices. Our future work will focus on evaluating our technique on multicore platform.
